\documentclass[aps,prd,preprintnumbers,showpacs,twocolumn,groupedaddress,nofootinbib]{revtex4}
\usepackage{graphicx}
\usepackage{latexsym}
\def\beq{\begin{equation}}
\def\eeq{\end{equation}}
\def\bey{\begin{eqnarray}}
\def\eey{\end{eqnarray}}

\def\lsim{\mathrel{\raise.3ex\hbox{$<$\kern-.75em\lower1ex\hbox{$\sim$}}}}
\def\gsim{\mathrel{\raise.3ex\hbox{$>$\kern-.75em\lower1ex\hbox{$\sim$}}}}

\begin{document}

\title{Implications of a Large $B_s \rightarrow \mu^+ \mu^-$ Branching Fraction for the Minimal Supersymmetric Standard Model}  
\author{Dan Hooper$^{1,2}$ and Chris Kelso$^3$}
\affiliation{$^1$Center for Particle Astrophysics, Fermi National Accelerator Laboratory, Batavia, IL 60510, USA}
\affiliation{$^2$Department of Astronomy and Astrophysics, University of Chicago, Chicago, IL 60637, USA}
\affiliation{$^3$Department of Physics, University of Chicago, Chicago, IL 60637, USA}

\date{\today}

\begin{abstract}

Very recently, the CDF Collaboration reported the first non-zero measurement of the $B_s \rightarrow \mu^+ \mu^-$ branching fraction. The central value of this measurement is more than 5 times of that predicted in the Standard Model and, if confirmed, would indicate the existence of new physics. We consider the implications of this measurement for the specific case of the Minimal Supersymmetric Standard Model (MSSM), and find that it requires large values of $\tan \beta$ ($\gsim 30$) and favors moderate values for the masses of the heavy higgs bosons ($m_A, m_H \sim 300-1200$ GeV). We also discuss the implications of this measurement for neutralino dark matter, finding that (within the MSSM) regions of parameter space in which the lightest neutralino can efficiently annihilate through the pseudoscalar higgs resonance (the $A$-funnel region) are favored.

\end{abstract}

\pacs{13.20.He, 11.30.Pb, 14.80.Nb, 14.80.Da; FERMILAB-PUB-11-332-A}
\maketitle

The study of rare decay modes can provide valuable probes of physics beyond the Standard Model, not easily accessible by other means. Of particular interest are the leptonic decays of the $B_s$ ($s\bar{b}$, $\bar{s}b$) and $B_d$ ($d\bar{b}$, $\bar{d}b$) mesons. In the Standard Model, such decays are dominated by $Z$ penguin and box diagrams which include a top quark loop. As the amplitudes for these processes are helicity suppressed and thus proportional to the mass of the final state leptons, one might expect $B_s$ and $B_d$ decays to $\tau^+ \tau^-$ to be most easily measured. Searches involving tau leptons are very difficult at hadron colliders, however, making such decays currently experimentally inaccessible. For these and other reasons, the rare decay modes $B_s \rightarrow \mu^+ \mu^-$ and $B_d \rightarrow \mu^+ \mu^-$ are among the most promising channels with which to constrain or infer physics beyond the Standard Model. 

Over the past several years, the CDF~\cite{:2007kv} and D0~\cite{Abazov:2010fs} collaborations (and more recently, LHCb~\cite{Aaij:2011rj}) have reported increasingly stringent upper limits on the $B_s \rightarrow \mu^+ \mu^-$ branching fraction, steadily moving closer to the value predicted in the Standard Model,  $(3.2 \pm 0.2) \times 10^{-9}$~\cite{Buras:2010mh}. Very recently, the CDF collaboration reported the first measurement of the $B_s \rightarrow \mu^+ \mu^-$ branching fraction inconsistent with a value of zero (at the level of 2.8\,$\sigma$). Furthermore, CDF's measurement, $\mathcal{B}(B_s \rightarrow \mu^+ \mu^-)=1.8^{+1.1}_{-0.9}\times 10^{-8}$, favors a central value that is 5-6 times larger than predicted in the Standard Model. From this result, the CDF collaboration excludes the branching fraction predicted by the Standard Model at the 98.1\% confidence level (C.L.)~\cite{Collaboration:2011fi}.

It has long been appreciated that extensions of the Standard Model, including supersymmetry, can lead to large enhancements of the $B_s \rightarrow \mu^+ \mu^-$ branching fraction~\cite{Choudhury:1998ze,Babu:1999hn}. In particular, in supersymmetric models with large values of $\tan \beta$ (the ratio of the vacuum expectation values of the two higgs doublets), this branching fraction can be as large as $\sim$10-100 times the value predicted in the Standard Model~\cite{Kane:2003wg,Dedes:2004yc,Ellis:2005sc}. In this letter, we explore the implications of CDF's new measurement for supersymmetry, focusing for concreteness on the Minimal Supersymmetric Standard Model (MSSM).

\begin{figure*}[t]
\centering
%\vspace{-1.0cm}
\includegraphics[angle=0.0,width=2.362in]{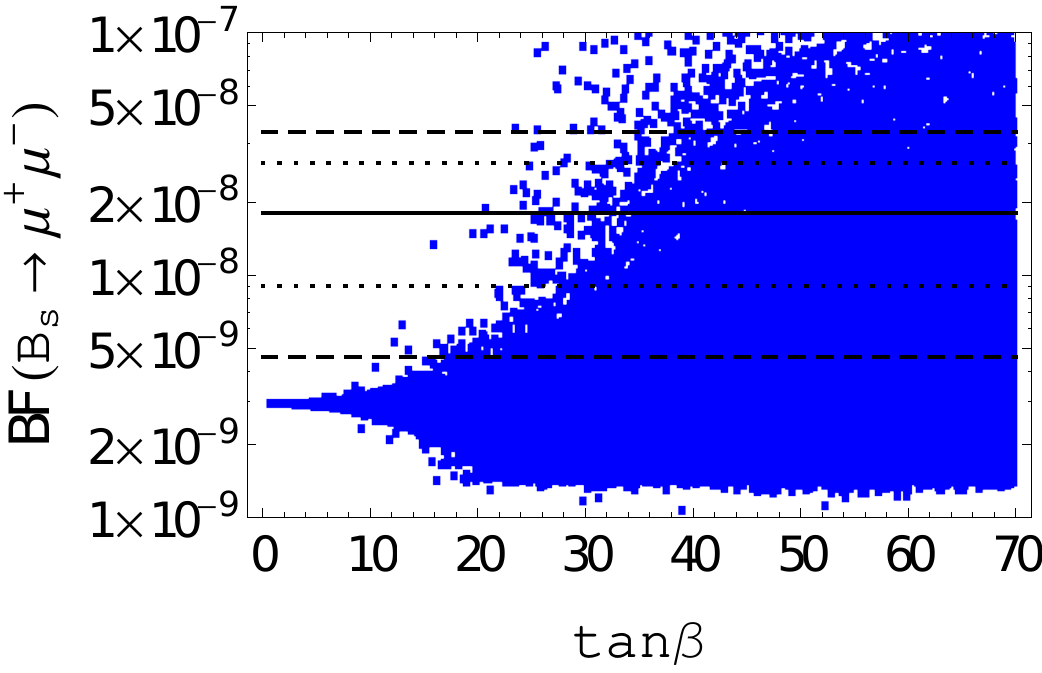}
\includegraphics[angle=0.0,width=2.24in]{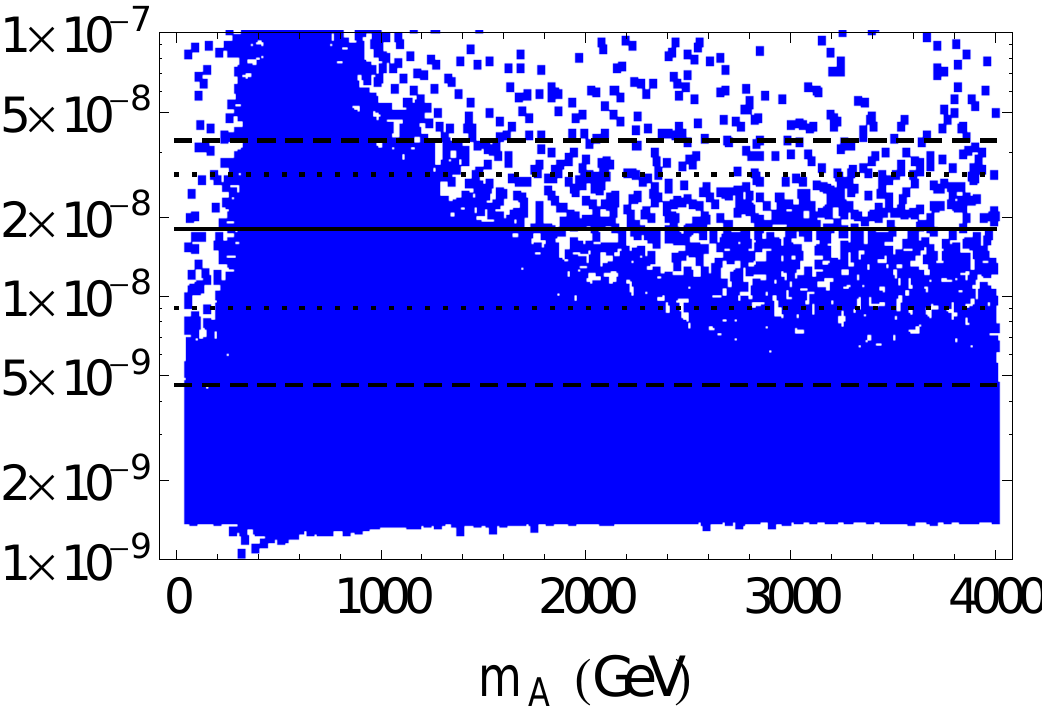}
\includegraphics[angle=0.0,width=2.19in]{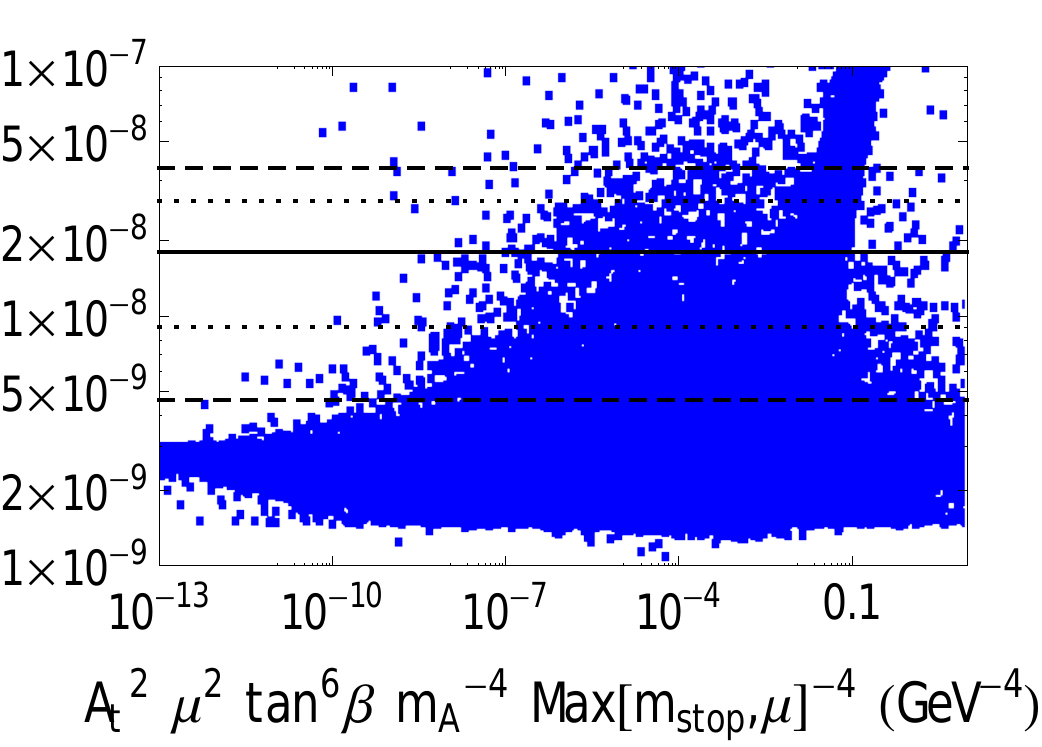}\\
\includegraphics[angle=0.0,width=2.362in]{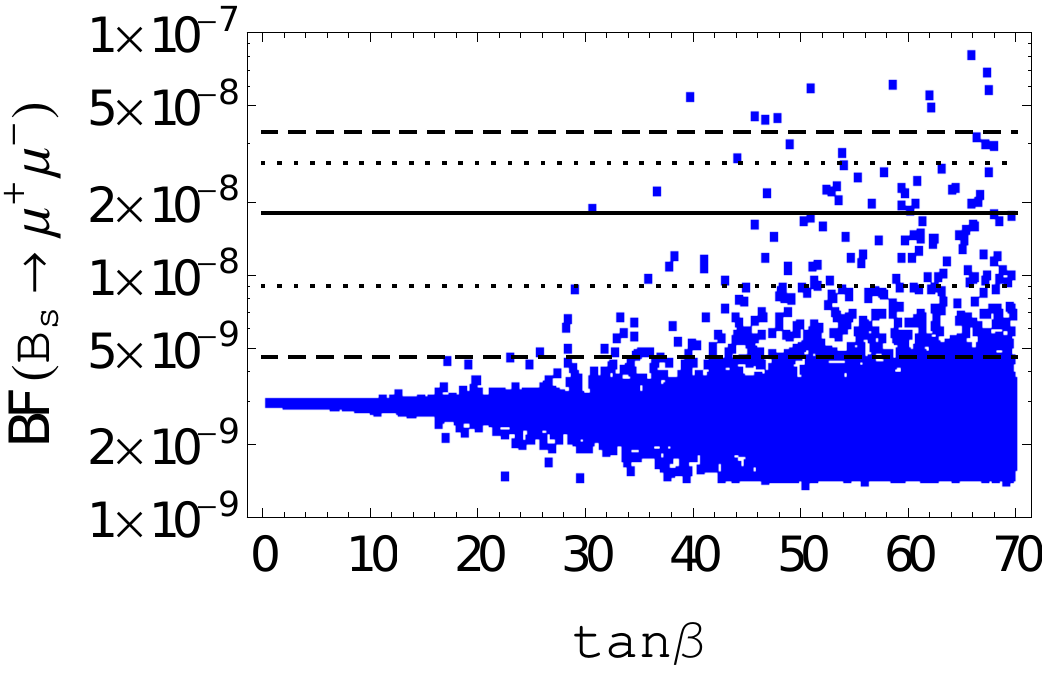}
\includegraphics[angle=0.0,width=2.24in]{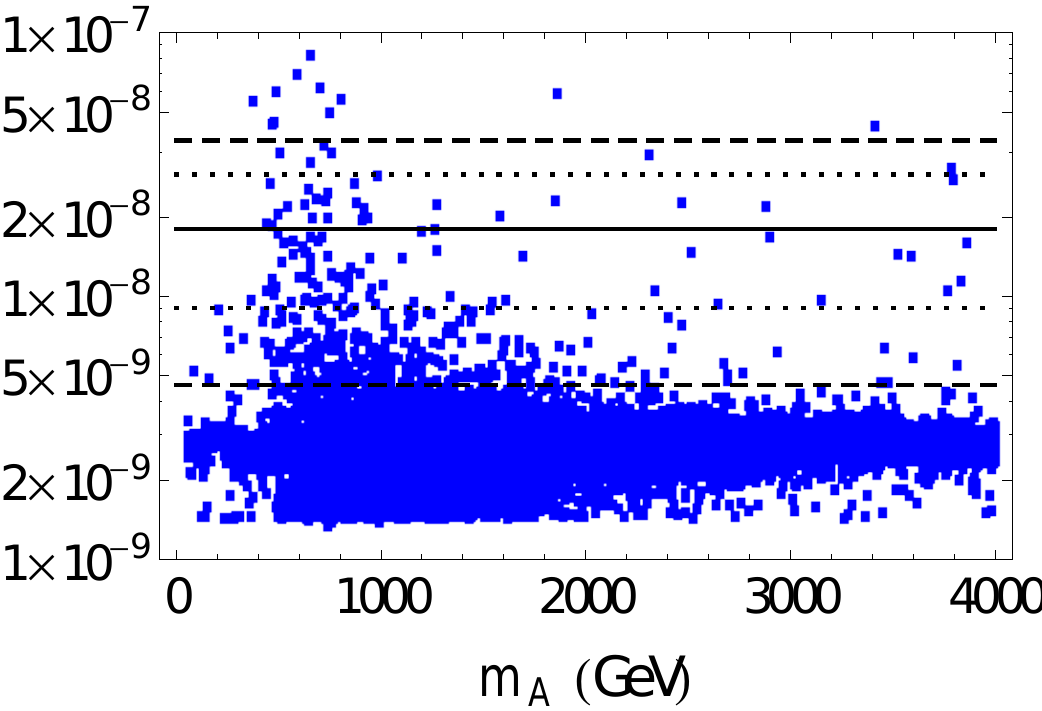}
\includegraphics[angle=0.0,width=2.17in]{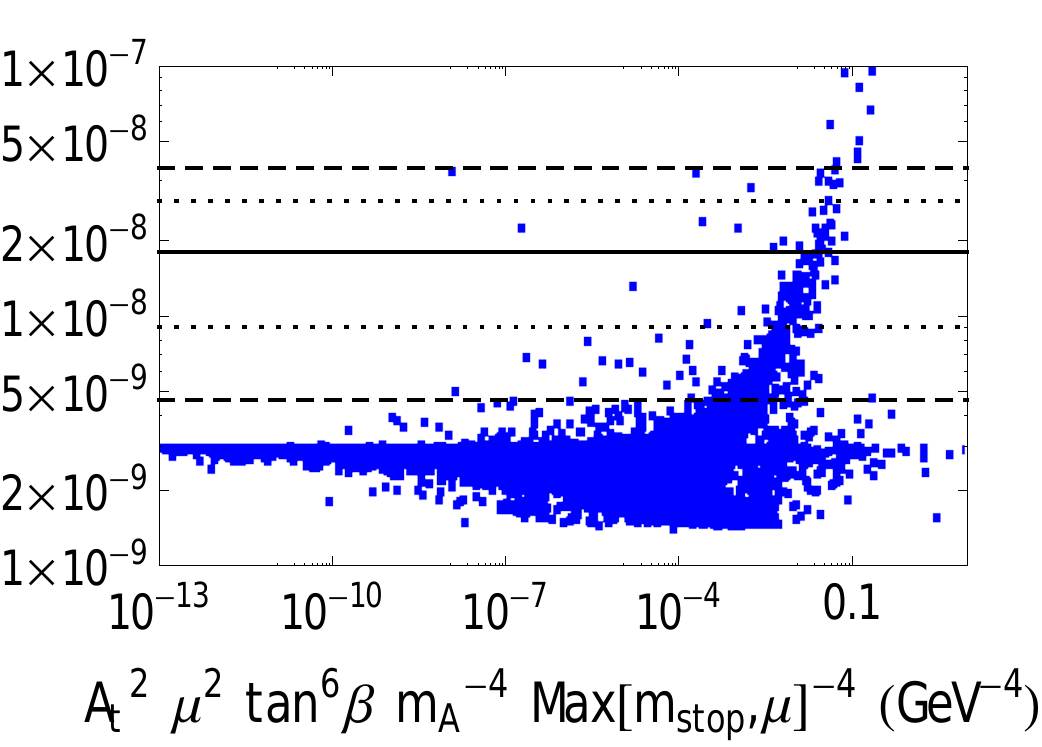}
\caption{Values of the $B_s \rightarrow \mu^+ \mu^-$ branching fraction over MSSM parameter space as a function of $\tan\beta$ (left), $m_A$ (center), and $A^2_t \, \mu^2 \, \tan^6\beta \, m^{-4}_A \, {\rm Max}(\mu, m_{\rm stop})^{-4}$ (right), where $m_{\rm{stop}}$ is the average of the two stop masses. In the lower frames, the thermal relic abundance of neutralino dark matter is required to fall within the measured range of the dark matter density (no such constraint was imposed in the upper frames). In each frame, we have applied the collider constraints as described in the text. The solid horizontal line denotes the central value as measured by CDF, whereas the dotted and dashed lines represent the 1$\sigma$ and 90\% C.L. ranges of this measurement, respectively.}
\label{scan}
\end{figure*}

The branching fraction for $B_s \rightarrow \mu^+ \mu^-$ can be written as~\cite{Bobeth:2001sq,Arnowitt:2002cq}:
\begin{eqnarray}
\mathcal{B}[B_s\rightarrow \mu^+ \mu^-] &=& \frac{\tau_B m^5_{B_s}}{32 \pi} f^2_{B_s} \sqrt{1-4m^2_{\mu}/m^2_{B_s}} \\
&\times&\bigg[\bigg(1-\frac{4m^2_{\mu}}{m^2_{B_s}}\bigg) \left|\frac{(C_S-C'_S)}{(m_b+m_s)}\right|^2 \nonumber \\
&+&\left|\frac{(C_P-C'_P)}{(m_b+m_s)}+2 \frac{m_{\mu}}{m^2_{B_s}}(C_A-C'_A)\right|^2\bigg] \nonumber,
\end{eqnarray}
where $f_{B_s}$ is the $B_s$ decay constant, and $m_{B_s}$ and $\tau_B$ are the mass and lifetime of the $B_s$ meson, respectively. The Wilson coefficients, $C_S$, $C'_S$, $C_P$ and $C'_P$, describe the relevant short distance physics, including any contributions from supersymmetric particles. To accommodate the large branching fraction measured by CDF, we focus on models with large $\tan \beta$. In this case, the process of $B_s\rightarrow \mu^+ \mu^-$ is dominated by diagrams with a higgs boson ($A$ or $H$) and a chargino-stop loop, leading to a contribution (neglecting QCD corrections) approximately given by:
\begin{eqnarray}
C_P &\approx& -C_S \approx \frac{-G_F \, \alpha}{\sqrt{2}\pi} V_{tb}V^*_{ts} \bigg(\frac{\tan^3\beta \, m_b \, m_{\mu} \, m_t \, \mu  \, \sin 2 \theta_{\tilde{t}}}{8 m^2_W\, m^2_A \, \sin^2\theta_W}\bigg)  \nonumber \\
\nonumber \\
&\times&\bigg(\frac{m^2_{\tilde{t}_1} \log[m^2_{\tilde{t}_1}/\mu^2]}{\mu^2-m^2_{\tilde{t}_1}}-\frac{m^2_{\tilde{t}_2} \log[m^2_{\tilde{t}_2}/\mu^2]}{\mu^2-m^2_{\tilde{t}_2}}\bigg),
\end{eqnarray}
where $m_{\tilde{t}_{1,2}}$ are the masses of the top squarks, $m_A$ is the mass of the pseudoscalar higgs, $\theta_{\tilde{t}}$ is the angle that diagonalizes the stop mass matrix, and $V_{tb}$, $V_{ts}$ are elements of the Cabibbo-Kobayashi-Maskawa (CKM) matrix. $C'_P$ and $C'_S$ are each suppressed by a factor of $m_s/m_b$ and thus provide only subdominant contributions. By inspection, we see that the dominant supersymmetric contributions to the $B_s \rightarrow \mu^+ \mu^-$ branching fraction scale with $\tan^6 \beta$, as well as with $m^{-4}_A$ from the propagator, $\mu^2 A^2_t$ from the mass insertions in the sparticle loop, and with either $\mu^{-4}$ or $m^{-4}_{\tilde{t}}$ (depending on which mass in the sparticle loop is heavier). 

To explore this quantitatively, we have used the program Micromegas~\cite{Belanger:2010pz} to calculate the $B_s \rightarrow \mu^+ \mu^-$ branching fraction over a large range of the MSSM parameter space. In scanning over supersymmetric parameters, we do not assume any particular supersymmetric breaking mechanism(s) but instead allow the parameters to be chosen independently from one another. We consider parameters over the following ranges: $M_1$ and slepton mass parameters up to 2 TeV, $\mu$ up to 3 TeV, $M_2$ and $m_A$ up to 4 TeV, squark and gluino mass parameters up to 10 TeV, $A_t$ up to 4 TeV (but not in excess of the stop masses) and $\tan \beta$ up to 70. For all mass parameters, we allow both positive and negative values. We require that the lightest supersymmetric particle be uncolored and electrically neutral, and impose constraints on higgs and charged sparticle masses from LEP-II. We also impose the recent constraints on squark and gluino masses from ATLAS (based on 165 pb$^{-1}$ of data)~\cite{squarksgluinos}, and constraints on the $\tan \beta$-$m_A$ plane from CMS (based on only 36 pb$^{-1}$)~\cite{Chatrchyan:2011nx}. Lastly, we impose that the $b\rightarrow \tau^{\pm} \nu$~\cite{Barlow:2011fu} and $b\rightarrow s \gamma$~\cite{PDG} branching fractions fall within $2\sigma$ of their measured values. At this point, we apply only the upper limit on the magnetic moment of the muon, but will return to this issue later in this letter.

In Fig.~\ref{scan}, we show the distribution of values of the $B_s \rightarrow \mu^+ \mu^-$ branching fraction (after imposing the previously described constraints) as a function of $\tan \beta$ (left), $m_A$ (center), and $A^2_t \, \mu^2 \, \tan^6\beta \, m^{-4}_A \, {\rm Max}(\mu, m_{\rm stop})^{-4}$ (right), where $m_{\rm stop}$ is the average of the two stop masses. The correlation between this branching fraction and $\tan \beta$ is particularly striking, essentially requiring large to moderate values of this quantity ($\tan \beta \gsim 30$). Although CDF's $B_s \rightarrow \mu^+ \mu^-$ branching fraction measurement also favors moderate or low values of $m_A$, a number of acceptable models were found with quite heavy $m_A$. (Note that although models with $m_A \lsim 300$ GeV and large $\tan \beta$ often lead to a large $B_s \rightarrow \mu^+ \mu^-$ branching fraction, this combination is inconsistent with the constraints from CMS's di-tau searches~\cite{Chatrchyan:2011nx}.)

\begin{figure}[t]
\centering
{\includegraphics[angle=0.0,width=2.75in]{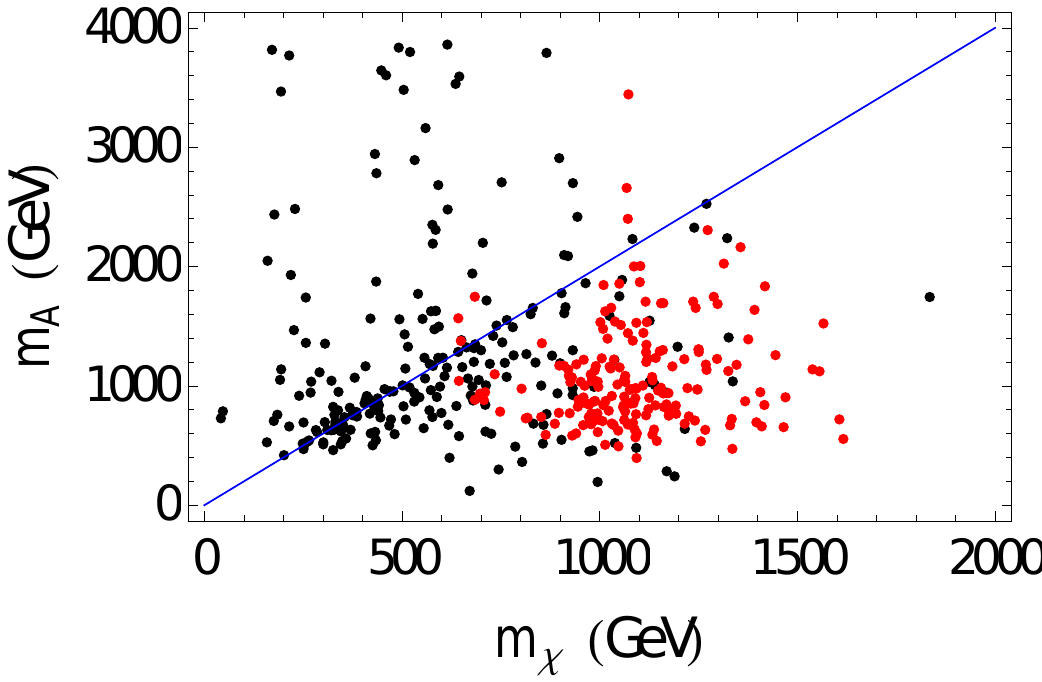}} \vspace{0.2cm}\\
\hspace{-0.3cm}
{\includegraphics[angle=0.0,width=2.88in]{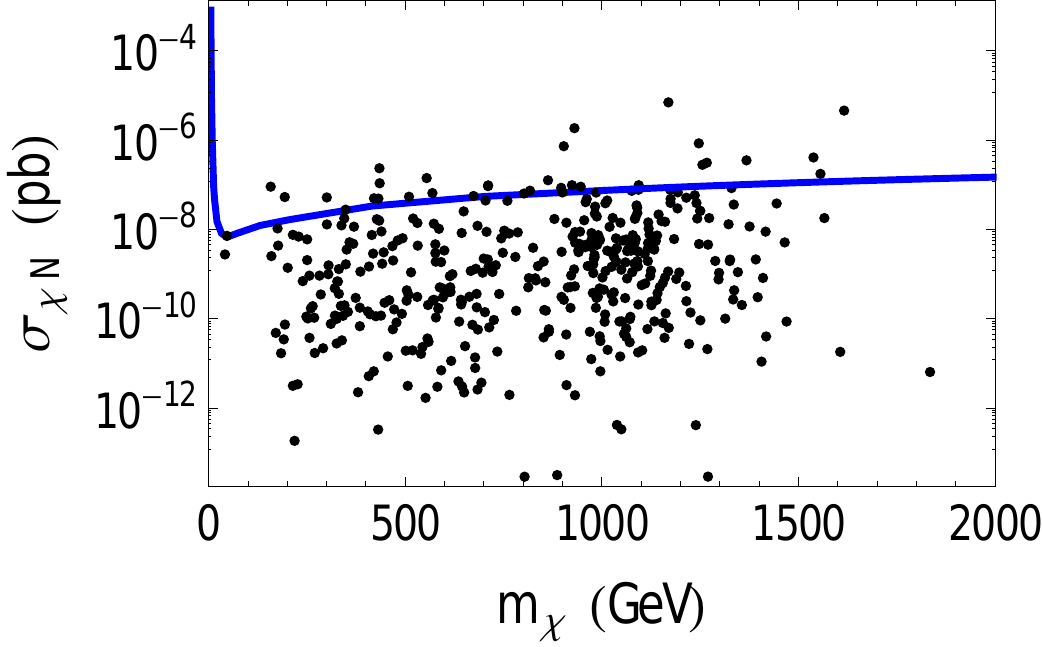}}
\caption{Implications of CDF's measurement of $\mathcal{B}(B_s \rightarrow \mu^+ \mu^-)$ for neutralino dark matter. In the upper frame, we compare the mass of the lightest neutralino to the mass of the pseudoscalar higgs in models which fall within the 90\% C.L. range of CDF's measurement, demonstrating that many of the models fall near the $A$-resonance ($2 m_{\chi} \approx m_A$). In this frame, black (red) points denote models with a bino-like neutralino (wino or higgsino-like neutralino). In the lower frame, we show the spin-independent elastic scattering cross section for neutralinos with nucleons, within this same set of models. Although the current constraint from XENON-100~\cite{xenon} excludes only a small fraction of these models, next generation direct detection experiments will be capable of testing a large fraction of the favored parameter space.}
\label{DM}
\end{figure}

In the lower three frames of Fig.~\ref{scan}, we also require that the lightest neutralino freezes-out in the early universe with a relic abundance within the range inferred by WMAP~($\Omega_{\rm CDM} h^2 = 0.1120 \pm 0.0056$)~\cite{Komatsu:2010fb}. This requirement removes a large fraction of the models found by our scan. In particular, the lightest neutralino is predicted to be overproduced in the early universe over much of supersymmetric parameter space. Regions of parameter space which do not suffer from this problem include those in which the lightest neutralino efficiently coannihilates with a nearly degenerate sparticle, efficiently annihilates due to its large couplings (made possible by sizable higgsino or wino components of its composition), or efficiently annihilates through the resonant exchange of the pseudoscalar higgs boson, $A$. As the cross section for neutralino annihilation to $b\bar{b}$ and $\tau^+ \tau^-$ through $A$-exchange is proportional to $\tan^2\beta$ and $1/m^4_A$, we expect the $A$-resonance to be highly efficient in many of the models favored by CDF's measurement of $\mathcal{B}[B_s \rightarrow \mu^+ \mu^-$]. In in upper frame of Fig.~\ref{DM}, we compare the masses of the lightest neutralino and the pseudoscalar higgs in models which yield $\mathcal{B}(B_s \rightarrow \mu^+ \mu^-)$ within the 90\% C.L. region as measured by CDF, and which predict a neutralino relic abundance consistent with WMAP. We find that many of the models in which the lightest neutralino is mostly bino-like (black points) fall near the $A$-resonance ($2 m_{\chi} \approx m_A$). Those models with wino-like or higgsino-like neutralinos (red points), however, can annihilate efficiently through chargino and/or neutralino exchange and thus need not (and often do not) lie near this contour. In the lower frame of Fig.~\ref{DM}, we plot the spin-independent elastic scattering cross section of the neutralino with nucleons, and compare this to the current upper limits from the XENON-100 experiment~\cite{xenon}. Although we find that direct detection experiments only rule out a small fraction of those models favored by CDF's measurement of $\mathcal{B}(B_s \rightarrow \mu^+ \mu^-)$, the elastic scattering cross sections predicted in many of these models are not far from the current constraints and are expected to fall within the reach of next generation experiments.

Next, we turn out attention to the anomalous magnetic moment of the muon, which has been measured to be $a^{\rm exp}_{\mu}=11 \,659 \,2080(63)\times 10^{-11}$~\cite{gminus2}.  When compared to the prediction of the Standard Model, $a^{\rm SM}_{\mu}=11\, 659\, 1790(65)\times 10^{-11}$~\cite{Jegerlehner:2009ry}, this measurement constitutes a 3.2$\sigma$ discrepancy, $\delta a_{\mu} \equiv a^{\rm exp}_{\mu}-a^{\rm SM}_{\mu} = (290 \pm 90)\times 10^{-11}$. To an extent, the supersymmetric contributions to $\delta a_{\mu}$ and $\mathcal{B}(B_s \rightarrow \mu^+ \mu^-)$ are correlated. In particular, as with the $B_s$ rare decay, large contributions to $\delta a_{\mu}$ are found only within the parameter space with large $\tan \beta$ and with light sparticles. As a concrete example, it has been previously shown that within the Constrained MSSM (CMSSM) with $M_{1/2}=450$ GeV, $M_{0}=350$ GeV, $A_{0}=0$, and $\mu >0$, the 1$\sigma$ range of $\delta a_{\mu}$ translates to a value of $\mathcal{B}(B_s \rightarrow \mu^+ \mu^-)$ that is no smaller than $1.0 \times 10^{-8}$~\cite{Dedes:2001fv}, illustrating that the measurements of $\delta a_{\mu}$ and $BR(B_s \rightarrow \mu^+ \mu^-)$ each favor overlapping parameter space with large values of $\tan \beta$. In more general terms, for the large values of $\tan \beta$ favored by CDF's measurement of $BR(B_s \rightarrow \mu^+ \mu^-)$, the observed value of $\delta a_{\mu}$ can be accommodated, so long as the masses of the contributing slepton and neutralino/chargino are not much heavier than a few hundred GeV.

In summary, we find that to accommodate the large $B_s \rightarrow \mu^+ \mu^-$ branching fraction reported by the CDF collaboration within the context of the Minimal Supersymmetric Standard Model (MSSM), one is forced to consider models with large values of $\tan \beta$ and (in most cases) modest values of $m_A$. Relatively large values of $A_t$ and light stops are also somewhat favored. In much of the supersymmetric parameter space favored by this measurement, neutralino dark matter annihilates efficiently through the resonant exchange of the pseudoscalar higgs, $A$, and scatters elastically with nuclei at a rate not far below current constraints from direct detection experiments.

In the immediate future, the LHC is expected to become sensitive to the $B_s \rightarrow \mu^+ \mu^-$ branching fraction as reported by CDF. Earlier this year, for example, the LHCb collaboration placed an upper limit of $5.6\times 10^{-8}$ (only a factor of three above CDF's central value) on this branching fraction using only 37 pb$^{-1}$ of data~\cite{Aaij:2011rj}. At present, the LHC experiments have each recorded approximately 30 times as much data, and should very quickly surpass CDF and D0 in sensitivity to $B_s \rightarrow \mu^+ \mu^-$. This will not only provide an opportunity to confirm CDF's measurement, but will also likely enable a more precise determination of this quantity.

\bigskip
\newpage

{\it Acknowledgements}: We would like to thank William Wester for valuable discussions. DH is supported by the US Department of Energy and by NASA grant NAG5-10842.

As we were completing this study, Refs.~\cite{new,new2} appeared which also discuss CDF's measurement of $B_s \rightarrow \mu^+ \mu^-$ within the context of supersymmetry.

\end{document}